# Radiative heat transfer in many-body systems: coupled electric and magnetic dipole approach

Jian Dong[a], Junming. Zhao[a*], Linhua Liu[a, b†]

[a] *School of Energy Science and Engineering, Harbin Institute of Technology, Harbin 150001, China*
[b] *Department of Physics, Harbin Institute of Technology, Harbin 150001, China*

**ABSTRACT**

The many-body radiative heat transfer theory [P. Ben-Abdallah, S.-A. Biehs, and K. Joulain, Phys. Rev. Lett. **107**, 114301 (2011)] only considered the contribution from the electric dipole moment. For metal particles, however, the magnetic dipole moment due to eddy current plays an important role, which can further couple with the electric dipole moment to introduce crossed terms. In this work, we develop coupled electric and magnetic dipole (CEMD) approach for the radiative heat transfer in a collection of objects in mutual interaction. Due to the coupled electric and magnetic interactions, four terms, namely the electric-electric, the electric-magnetic, the magnetic-electric and the magnetic-magnetic terms, contribute to the radiative heat flux and the local energy density. The CEMD is applied to study the radiative heat transfer between various dimers of nanoparticles. It is found that each of the four terms can dominate the radiative heat transfer depending on the position and composition of particles. Moreover, near-field many-body interactions are studied by CEMD considering both dielectric and metallic nanoparticles. The near-field radiative heat flux and local energy density can be greatly increased when the particles are in coupled resonances. Surface plasmon polariton and surface phonon polariton can be coupled to enhance the radiative heat flux.

## I. INTRODUCTION

Radiative transfer at nanoscale attracts the interests of many researchers due to its significance in a wide range of scientific and engineering disciplines [1,2]. Near-field radiative heat transfer (NFRHT), in particular, can largely exceed the black body limit due to the tunneling of evanescent waves [3-6], which cannot be predicted by the classical radiative transfer theory [2]. With such a remarkable feature, NFRHT plays an important role in the study of nanoscale heat transfer and finds promising applications in energy conversion[7-11], thermal rectification [12-15], thermal photon based logic [16,17] and memory [18], etc. Over the past few years, many theoretical approaches on NFRHT problems have been put forward by combining the Maxwell electromagnetic theory and the fluctuation-dissipation theorem [3]. These approaches, including the Green's function [3,19-21], the scattering matrix [22-26], the finite difference time domain [27-31], the thermal discrete dipole approximation [32-34], the rigorous coupled wave analysis [35-38], the fluctuating surface [39-41] and volume [42-44] current etc., greatly enrich our understanding of NFRHT problems. Meanwhile, more and more experimental researches on NFRHT have been performed [45-53].

[*] jmzhao@hit.edu.cn (Junming Zhao)
[†] lhliu@hit.edu.cn (Linhua Liu)

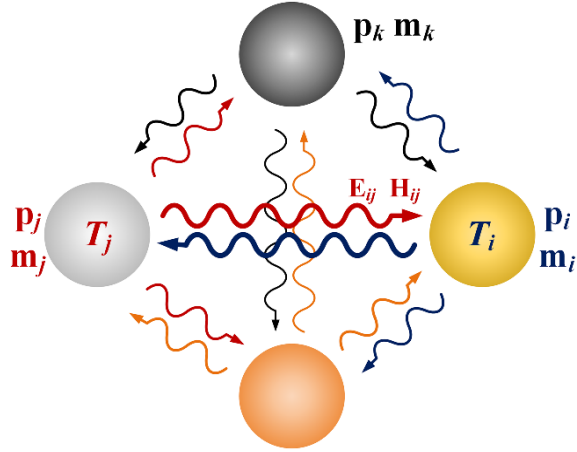

FIG. 1. Schematic of the many-body radiative heat transfer.

While most of the studies focused on NFRHT between two objects, a few recent theoretical studies have been conducted for more than two objects, with an emphasis on the many-body interactions in the near-field. Ben-Abdallah et al. [54] developed the many-body radiative heat transfer theory in the framework of the coupled electric dipole method and observed a strong exaltation of radiative heat flux between SiC nanoparticles due to the many-body interactions. With the aid of this theory, a heat superdiffusion phenomenon was demonstrated in a SiC nanoparticle system [55]. Employing the many-body theory, Phan et al. [56] studied the NFRHT between gold nanoparticle arrays where they considered the electric and magnetic dipole moments but treated them separately. Dong et al. [57] applied the many-body theory to investigate the radiative heat transfer between two clusters of SiC nanoparticles and found that the many-body interactions in the clusters inhibited the radiative heat transfer. This theory has also been extended to anisotropic electric polarizabilities and it was shown that the NFRHT can be sensitively altered by the shape and the relative orientation of the particles [58,59]. Recently, Ben-Abdallah [60] reported the photon thermal hall effect in a network of magneto-optical particles subjected to a constant magnetic field. In addition, Messina et al. [61] put forward a general fluctuation-electrodynamic theory to study the dynamics of heat transfer in nanoparticle systems. They found that near-field many-body interactions can dramatically tailor the temperature field distribution and change the thermal relaxation process. The dynamic many-body NFRHT has also been considered in three anisotropic particles systems [62] and in the heating of a collection of nanoparticles by incident laser pulses [63,64].

However, the many-body radiative heat transfer theories mentioned above did not include the mutual interactions of the electric and magnetic dipole moments. Most of the studies considered dielectric particles for which the magnetic dipole moment was neglected. This is reasonable for non-magnetic dielectric nanoparticles since the magnetic polarizability (being proportional to $R^5/\lambda^2$ where $R$ is particle radius and $\lambda$ the wavelength) is much smaller than the electric one (being proportional to $R^3$) [65]. For metallic particles, however, the magnetic dipole moment due to eddy currents can dominate the NFRHT [21,66]. Moreover, for a system containing both dielectric and metallic particles, the electric and magnetic crossed terms may play an important role as shown by Manjavacas and de Abajo [67] for two particle problems. Besides, although some of the numerical methods (scattering matrix, finite difference time domain, fluctuating volume current for instance) can in principle deal with many-body problems, they are computationally expensive and usually applied to problems of two objects.

In this work, therefore, we develop a coupled electric and magnetic dipole (CEMD) approach for the radiative heat transfer in many-body systems. The CEMD was initially introduced to deal with light scattering problems of agglomerates [65]. In CEMD, each particle is assigned with an electric and a magnetic dipole moment, and then the electric and magnetic dipole moments are solved considering the mutual interactions among the nanoparticles. CEMD is simple and efficient, which has been frequently applied to the light scattering problems of nanoparticle arrays [68-70]. This work is organized as follows. In Sec. II, we present the theoretical aspects of the CEMD for the radiative heat transfer in many-body systems. A validation of CEMD is given by comparing to the exact method for two sphere problems. In Sec. III, the CEMD is applied to calculate radiative heat flux between different dimers of nanoparticles. Then the effects of the near-field many-body interactions on the NFRHT and the local energy density distribution are studied considering both dielectric and metallic nanoparticles.

## II. THEORETICAL ASPECTS

We consider a set of $N$ nanoparticles located in a non-absorbing environment. The sizes of the nanoparticles are supposed to be much smaller than the characteristic thermal wavelength $\lambda_T = \hbar c / k_B T$ so that the nanoparticles can be treated as electric and magnetic dipoles in mutual interactions. All the materials considered are assumed non-magnetic and the magnetic dipole moments are induced by the eddy currents. The electromagnetic fields are those generated by the fluctuating electric and magnetic dipole moments of the nanoparticles. All the equations and formulas are written in Gaussian unit in this section.

### A. Green's functions of the electric and magnetic field in a nanoparticle system

We first consider the electromagnetic fields induced by an electric and magnetic dipole moment. The electric $\mathbf{E}_p$ and magnetic $\mathbf{H}_p$ fields at position $\mathbf{r}_i$ from an electric dipole moment $\mathbf{p}$ at $\mathbf{r}_j$ can be expressed in the following form [71]

$$\mathbf{E}_p = k^2 (\hat{\mathbf{r}} \times \mathbf{p}) \times \hat{\mathbf{r}} \frac{e^{ikr}}{r} + \left[ 3\hat{\mathbf{r}}(\hat{\mathbf{r}} \cdot \mathbf{p}) - \mathbf{p} \right] \left( \frac{1}{r^3} - \frac{ik}{r^2} \right) e^{ikr} \quad (1)$$

$$\mathbf{H}_p = k^2 (\hat{\mathbf{r}} \times \mathbf{p}) \frac{e^{ikr}}{r} \left( 1 - \frac{1}{ikr} \right) \quad (2)$$

where $k$ is the wave vector, $r$ is the magnitude of the vector $\mathbf{r} = \mathbf{r}_i - \mathbf{r}_j$, and $\hat{\mathbf{r}} = \mathbf{r}/r$ is the unit vector of $\mathbf{r}$. Similarly, the corresponding expressions for the electric $\mathbf{E}_m$ and magnetic $\mathbf{H}_m$ fields at position $\mathbf{r}_i$ from an magnetic dipole moment $\mathbf{m}$ at $\mathbf{r}_j$ are given by [71]

$$\mathbf{H}_m = k^2 (\hat{\mathbf{r}} \times \mathbf{m}) \times \hat{\mathbf{r}} \frac{e^{ikr}}{r} + \left[ 3\hat{\mathbf{r}}(\hat{\mathbf{r}} \cdot \mathbf{m}) - \mathbf{m} \right] \left( \frac{1}{r^3} - \frac{ik}{r^2} \right) e^{ikr} \quad (3)$$

$$\mathbf{E}_m = -k^2 (\hat{\mathbf{r}} \times \mathbf{m}) \frac{e^{ikr}}{r} \left( 1 - \frac{1}{ikr} \right) \quad (4)$$

respectively. For ease of analysis, the electric and magnetic fields and dipole moments are combined as

$$\mathbb{E} = \begin{pmatrix} \mathbf{E} \\ \mathbf{H} \end{pmatrix} , \quad \mathbb{P} = \begin{pmatrix} \mathbf{p} \\ \mathbf{m} \end{pmatrix} \quad (5)$$

where $\mathbb{E}$ and $\mathbb{P}$ are 6×1 vectors. Then the electric and magnetic field $\mathbb{E}$ at the place $\mathbf{r}_i$ from an electric and magnetic dipole moment $\mathbb{P}$ located at $\mathbf{r}_j$ can be casted in a more compact form as

$$\mathbb{E} = \mathbb{G}_{0,ij} \mathbb{P} \quad (6)$$

Here $\mathbb{G}_{0,ij} = \mathbb{G}_0(\mathbf{r}_i, \mathbf{r}_j, \omega)$ is the free space Green's function for the electric and magnetic field $\mathbb{E}$, which can be divided into four sub-terms

$$\mathbb{G}_{0,ij} = \begin{bmatrix} \hat{G}_{0,ij}^{EE} & \hat{G}_{0,ij}^{EM} \\ \hat{G}_{0,ij}^{ME} & \hat{G}_{0,ij}^{MM} \end{bmatrix} \quad (7)$$

Each of the four sub-terms is a 3×3 tensor. According to Eqs. (1) - (4), the sub-terms are given by [65]:

$$\hat{G}_{0,ij}^{EE} = \hat{G}_{0,ij}^{MM} = \frac{e^{ikr}}{r} \left[ \left( k^2 - \frac{1}{r^2} + \frac{ik}{r} \right) \mathbb{I} + \left( -k^2 + \frac{3}{r^2} - \frac{3ik}{r} \right) \hat{\mathbf{r}} \otimes \hat{\mathbf{r}} \right] \quad (8)$$

$$\hat{G}_{0,ij}^{EM} = -\hat{G}_{0,ij}^{ME} = -\frac{e^{ikr}}{r}\left(k^2 + \frac{ik}{r}\right)\begin{bmatrix} 0 & -\hat{\mathbf{r}}_z & \hat{\mathbf{r}}_y \\ \hat{\mathbf{r}}_z & 0 & -\hat{\mathbf{r}}_x \\ -\hat{\mathbf{r}}_y & \hat{\mathbf{r}}_x & 0 \end{bmatrix} \quad (9)$$

where $\mathbb{I}$ is the unit tensor, $r$ is the magnitude of the vector $\mathbf{r} = \mathbf{r}_i - \mathbf{r}_j$, $\hat{\mathbf{r}} = \mathbf{r}/r$ is the unit vector of $\mathbf{r}$, $\hat{\mathbf{r}}_{\nu=x,y,z}$ denotes the three components of the unit vector $\hat{\mathbf{r}}$, and $\otimes$ represents the outer product of vectors.

For an incident electric and magnetic field on the dipole, the induced electric and magnetic dipole moments are

$$\mathbf{p} = \alpha^E \mathbf{E} \quad , \quad \mathbf{m} = \alpha^M \mathbf{H} \quad (10)$$

where $\alpha^E$ and $\alpha^M$ are the electric and magnetic polarizabilities of the dipole, respectively. We here limit our discussions to isotropic and spherical particles. Hence, the polarizabilities are isotropic. The induced electric and magnetic moments $\mathbb{P}^{ind}$ by the incident field $\mathbb{E}^{inc}$ can be written in compact form by introducing the combined polarizability tensor $\hat{\alpha}$

$$\mathbb{P}^{ind} = \hat{\alpha}\mathbb{E}^{inc} = \begin{pmatrix} \alpha^E \mathbb{I}_3 & 0 \\ 0 & \alpha^M \mathbb{I}_3 \end{pmatrix}\mathbb{E}^{inc} \quad (11)$$

In a nanoparticle system, the mutual interactions must be taken into account. The dipole moment of the $i$-th particle is the sum of the part from the thermal fluctuations of the $i$-th particle $\mathbb{P}_i^{fluc}$ and the part induced by the fluctuating incident fields from other nanoparticles $\mathbb{P}_i^{ind}$

$$\mathbb{P}_i = \mathbb{P}_i^{fluc} + \mathbb{P}_i^{ind} \quad (12)$$

The induced dipole moment, according to the coupled electric and magnetic dipole method [65], can be written as

$$\mathbb{P}_i^{ind} = \hat{\alpha}_i \sum_{j \neq i} \mathbb{G}_{0,ij} \mathbb{P}_j \quad (13)$$

Taking Eq. (13) into Eq. (12), the dipole moment of each particle can then be organized in a matrix form as [54,61]

$$\begin{pmatrix} \mathbb{P}_1 \\ \vdots \\ \mathbb{P}_N \end{pmatrix} = \mathbb{A}^{-1} \begin{pmatrix} \mathbb{P}_1^{fluc} \\ \vdots \\ \mathbb{P}_N^{fluc} \end{pmatrix} \quad (14)$$

where the interaction matrix $\mathbb{A}$ is given by

$$\mathbb{A} = \mathbb{I} - \begin{bmatrix} 0 & \hat{\alpha}_1 \mathbb{G}_{0,12} & \cdots & \hat{\alpha}_1 \mathbb{G}_{0,1N} \\ \hat{\alpha}_2 \mathbb{G}_{0,21} & \ddots & \ddots & \vdots \\ \vdots & \ddots & \ddots & \hat{\alpha}_{(N-1)} \mathbb{G}_{0,(N-1)N} \\ \hat{\alpha}_N \mathbb{G}_{0,N1} & \cdots & \hat{\alpha}_N \mathbb{G}_{0,N(N-1)} & 0 \end{bmatrix} \quad (15)$$

Once the dipole moments are obtained, the corresponding radiating electric and magnetic fields by the particle system can be calculated according to Eq. (6). The electric and magnetic fields at position $\mathbf{r}_i$ due to the fluctuating dipole moments of the *j*-th particle at $\mathbf{r}_j$ can be written as

$$\mathbb{E}_{ij} = \mathbb{G}_{ij} \mathbb{P}_j^{fluc} \quad (16)$$

where $\mathbb{G}_{ij} = \mathbb{G}(\mathbf{r}_i, \mathbf{r}_j, \omega)$ is the Green's function (i.e. the propagator) in a nanoparticle system linking the dipole moments and the fields [54]. It can be calculated by [61]

$$\begin{pmatrix} \mathbb{G}_{i1} & \cdots & \mathbb{G}_{iN} \end{pmatrix} = \begin{pmatrix} \mathbb{G}_{0,i1} & \cdots & \mathbb{G}_{0,iN} \end{pmatrix} \mathbb{A}^{-1} \quad (17)$$

Similarly, the Green's function $\mathbb{G}_{ij}$ can be divided into four sub-terms

$$\mathbb{G}_{ij} = \begin{bmatrix} \hat{G}_{ij}^{EE} & \hat{G}_{ij}^{EM} \\ \hat{G}_{ij}^{ME} & \hat{G}_{ij}^{MM} \end{bmatrix} \quad (18)$$

Note that the Green's function $\mathbb{G}_{ij}$ differs from the free space Green's function $\mathbb{G}_{0,ij}$ in that it takes into account the mutual interactions of the nanoparticles. $\hat{G}_{ij}^{EE}$ and $\hat{G}_{ij}^{EM}$ link the electric field at position $\mathbf{r}_i$ to the electric and magnetic dipole moments of the *j*-th particle, respectively. Similarly, $\hat{G}_{ij}^{ME}$ and $\hat{G}_{ij}^{MM}$ link the magnetic field at position $\mathbf{r}_i$ to the electric and magnetic dipole moments of the *j*-th particle, respectively.

**B. Radiative heat flux and energy density**

For an incident electromagnetic wave on a particle, the power of the wave is attenuated due to absorption and scattering. The sum of the absorption and the scattering is the extinction, which for a nanoparticle with an electric and magnetic dipole moment can be calculated by [68]

$$P_{ext} = \omega \operatorname{Im}\left(\left\langle \mathbf{p}^{ind} \cdot \mathbf{E}^{inc*} \right\rangle + \left\langle \mathbf{m}^{ind} \cdot \mathbf{H}^{inc*} \right\rangle\right) \quad (19)$$

where $\mathbf{E}^{inc}$ and $\mathbf{H}^{inc}$ are the incident electromagnetic fields, $\mathbf{p}^{ind}$ and $\mathbf{m}^{ind}$ are the induced electric and magnetic dipole moments, respectively. The scattered power of an electric and magnetic dipole is [71]

$$P_{sca} = \frac{2}{3}\omega k^3 \left( \left|\mathbf{p}^{ind}\right|^2 + \left|\mathbf{m}^{ind}\right|^2 \right) \tag{20}$$

Thus the power absorbed by the *i*-th particle due to the incident electric and magnetic fields of the *j*-th particle is

$$P_{ij} = 2\int_0^{+\infty} \omega \frac{d\omega}{4\pi^2} \left[ \text{Im}\left(\left\langle \mathbf{p}_i^{ind} \cdot \mathbf{E}_{ij}^* \right\rangle + \left\langle \mathbf{m}_i^{ind} \cdot \mathbf{H}_{ij}^* \right\rangle \right) - \frac{2}{3}k^3 \left(\left\langle \left|\mathbf{p}_i^{ind}\right|^2 \right\rangle + \left\langle \left|\mathbf{m}_i^{ind}\right|^2 \right\rangle \right) \right] \tag{21}$$

In Eq. (21) we use the convention $f(t) = \int \frac{d\omega}{2\pi} f(\omega) e^{-i\omega t}$ for the time Fourier transform and consider only positive frequencies. Introducing the parameter $\chi_j^\nu = \alpha_j^\nu - \left(2k^3 i/3\right)\left|\alpha_j^\nu\right|^2$ ($\nu = E, M$), Eq. (21) can be written as

$$P_{ij} = \int_0^{+\infty} 2\omega \frac{d\omega}{4\pi^2} \left[ \text{Im}\left(\chi_i^E\right) \left\langle \mathbf{E}_{ij} \cdot \mathbf{E}_{ij}^* \right\rangle + \text{Im}\left(\chi_i^M\right) \left\langle \mathbf{H}_{ij} \cdot \mathbf{H}_{ij}^* \right\rangle \right] \tag{22}$$

According to Eq. (16) and (18), the incident electric and magnetic fields on the *i*-th particle due to the fluctuating dipole moments of the *j*-th particle are

$$\mathbf{E}_{ij} = \hat{G}_{ij}^{EE} \mathbf{p}_j^{fluc} + \hat{G}_{ij}^{EM} \mathbf{m}_j^{fluc} \tag{23}$$

$$\mathbf{H}_{ij} = \hat{G}_{ij}^{ME} \mathbf{p}_j^{fluc} + \hat{G}_{ij}^{MM} \mathbf{m}_j^{fluc} \tag{24}$$

In addition, the fluctuation dissipation theorem for the electric and magnetic dipole moment (in Gaussian units) reads [67]

$$\left\langle p_{j,\beta}^{fluc}(\omega) p_{j',\beta'}^{fluc*}(\omega') \right\rangle = \frac{4\pi}{\omega} \Theta(\omega, T) \text{Im}\left[\chi_{j,\beta\beta'}^E(\omega)\right] \delta(\omega - \omega') \delta_{jj'} \delta_{\beta\beta'} \tag{25}$$

$$\left\langle m_{j,\beta}^{fluc}(\omega) m_{j',\beta'}^{fluc*}(\omega') \right\rangle = \frac{4\pi}{\omega} \Theta(\omega, T) \text{Im}\left[\chi_{j,\beta\beta'}^M(\omega)\right] \delta(\omega - \omega') \delta_{jj'} \delta_{\beta\beta'} \tag{26}$$

where $\Theta(\omega, T) = \hbar\omega / \left(e^{\hbar\omega/k_B T} - 1\right)$ is the mean energy of a harmonic oscillator at equilibrium, $\hbar$ denotes the reduced Planck constant, $k_B$ denotes the Boltzmann's constant, *T* is the temperature. Notice that the zero point energy $\hbar\omega/2$ is not considered.

Taking Eq. (23) - (26) into Eq. (22) and assuming no correlation between the fluctuating electric and magnetic dipole moments, the power absorbed by the *i*-th particle due to the incident field of the *j*-th particle can be calculated by

$$P_{ij} = \int_0^{+\infty} \frac{d\omega}{2\pi} P_{ij,\omega} = 3\int_0^{+\infty} \frac{d\omega}{2\pi} \Theta(\omega, T_j) \mathcal{T}_{ij}(\omega) \tag{27}$$

where $\mathcal{T}_{ij}(\omega)$ is the transmission coefficient

$$\mathcal{T}_{ij}(\omega) = \frac{4}{3}\text{Im}(\chi_i^E)\text{Im}(\chi_j^E)\text{Tr}(\hat{G}_{ij}^{EE}\hat{G}_{ij}^{EE*}) + \frac{4}{3}\text{Im}(\chi_i^E)\text{Im}(\chi_j^M)\text{Tr}(\hat{G}_{ij}^{EM}\hat{G}_{ij}^{EM*}) + \\ \frac{4}{3}\text{Im}(\chi_i^M)\text{Im}(\chi_j^E)\text{Tr}(\hat{G}_{ij}^{ME}\hat{G}_{ij}^{ME*}) + \frac{4}{3}\text{Im}(\chi_i^M)\text{Im}(\chi_j^M)\text{Tr}(\hat{G}_{ij}^{MM}\hat{G}_{ij}^{MM*}) \tag{28}$$

which is similar to that in Ref. [54], but now four terms are included in the transmission coefficient, namely the electric-electric (EE) term $\mathcal{T}_{ij}^{EE}(\omega)$, the electric-magnetic (EM) term $\mathcal{T}_{ij}^{EM}(\omega)$, the magnetic-electric (ME) term $\mathcal{T}_{ij}^{ME}(\omega)$ and the magnetic-magnetic (MM) term $\mathcal{T}_{ij}^{MM}(\omega)$. The EM and ME terms are the crossed terms discussed in Ref. [67] for two particle problems. If the magnetic dipole moment is neglected, only $\mathcal{T}_{ij}^{EE}(\omega)$ remains, which is the same to the transmission coefficient in Ref. [54].

The energy density at position $\mathbf{r}_0$ of the radiating electromagnetic field from the *j*-th particle is

$$u_{\mathbf{r}_0 j} = \int_0^{+\infty} \frac{d\omega}{2\pi} u_{\mathbf{r}_0 j}(\omega) = 2\int_0^{+\infty} \frac{d\omega}{4\pi^2} \frac{1}{8\pi} \left( \langle \mathbf{E}_{\mathbf{r}_0 j} \cdot \mathbf{E}^*_{\mathbf{r}_0 j} \rangle + \langle \mathbf{H}_{\mathbf{r}_0 j} \cdot \mathbf{H}^*_{\mathbf{r}_0 j} \rangle \right) \tag{29}$$

where $u_{\mathbf{r}_0 j}(\omega)$ is the spectral local energy density. Considering the fluctuation dissipation theorem Eq. (25), (26) and the Green's functions Eq. (16), (18) in the particle system, $u_{\mathbf{r}_0 j}(\omega)$ can be written as [19]

$$u_{\mathbf{r}_0 j}(\omega) = \rho_{\mathbf{r}_0 j}(\omega) \Theta(\omega, T_j) \tag{30}$$

where $\rho_{\mathbf{r}_0 j}(\omega)$ is the local density of states (LDOS) [19,72] which is calculated by

$$\rho_{\mathbf{r}_0 j}(\omega) = \frac{1}{2\pi\omega} \begin{bmatrix} \text{Im}(\chi_j^E)\text{Tr}(\hat{G}_{\mathbf{r}_0 j}^{EE}\hat{G}_{\mathbf{r}_0 j}^{EE*}) + \text{Im}(\chi_j^M)\text{Tr}(\hat{G}_{\mathbf{r}_0 j}^{EM}\hat{G}_{\mathbf{r}_0 j}^{EM*}) + \\ \text{Im}(\chi_j^E)\text{Tr}(\hat{G}_{\mathbf{r}_0 j}^{ME}\hat{G}_{\mathbf{r}_0 j}^{ME*}) + \text{Im}(\chi_j^M)\text{Tr}(\hat{G}_{\mathbf{r}_0 j}^{MM}\hat{G}_{\mathbf{r}_0 j}^{MM*}) \end{bmatrix} \tag{31}$$

Like the transfer coefficient, there are also four terms contained in the local density of states.

## C. Polarizabilities of nanoparticles

For isotropic spherical particles, the electric and magnetic polarizabilities can be obtained from the extinction cross section of small spherical particles [65,73]

$$\alpha^E = (3i/2k^3)a_1 \ , \ \alpha^M = (3i/2k^3)b_1 \tag{32}$$

where $a_1$ and $b_1$ are the first order of the Mie coefficients. For isotropic, nonmagnetic spherical particles, they are given by [73]

$$a_n = \frac{\varepsilon j_n(y)[xj_n(x)]' - j_n(x)[yj_n(y)]'}{\varepsilon j_n(y)[xh_n^{(1)}(x)]' - h_n^{(1)}(x)[yj_n(y)]'} \quad (33)$$

$$b_n = \frac{j_n(y)[xj_n(x)]' - j_n(x)[yj_n(y)]'}{j_n(y)[xh_n^{(1)}(x)]' - h_n^{(1)}(x)[yj_n(y)]'} \quad (34)$$

where $x = kR$ and $y = \sqrt{\varepsilon}kR$, $R$ denotes the radius of the particle, and $\varepsilon$ denotes the dielectric permittivity of the particle, $j_n$ and $h_n^{(1)}$ are the Bessel functions and the Hankel functions of the first kind, respectively [73].

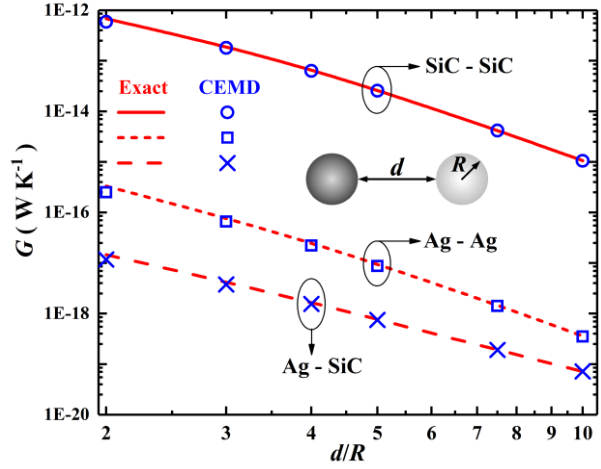

FIG. 2. The thermal conductance $G$ [see Eq. (35)] between two spherical nanoparticles as a function of $d/R$: comparison between the coupled electric and magnetic dipole (CEMD) method (symbols) and the exact method (lines) [20]. $d$ is the separation distance edge to edge between the spherical particles, the radius of the particles is $R$=50nm. The temperature is 300K. Three cases are considered, namely two SiC particles, two Ag particles, Ag and SiC particles.

As a validation of the approach, CEMD is compared to the exact method for two sphere problem proposed by Narayanaswamy and Chen [20]. The total conductance is defined as

$$G = \lim_{\Delta T \to 0} \Delta P / \Delta T \quad (35)$$

The exact method expands the field and the dyadic Green's function into vector spherical harmonics, in which higher multipoles are included. Three cases are considered, namely between two SiC nanoparticles, between two Ag nanoparticles, and between Ag and SiC nanoparticles. The dielectric function of SiC is described by the Drude-Lorentz's permittivity model [74]

$$\varepsilon(\omega) = \varepsilon_\infty \frac{\omega^2 - \omega_l^2 + i\Gamma\omega}{\omega^2 - \omega_t^2 + i\Gamma\omega} \quad (36)$$

where $\varepsilon_\infty = 6.7$, $\omega_l = 1.827 \times 10^{14}$ rad s$^{-1}$, $\omega_t = 1.495 \times 10^{14}$ rad s$^{-1}$, $\Gamma = 0.9 \times 10^{12}$ rad s$^{-1}$. The dielectric response of Ag is described by the Drude model [75]

$$\varepsilon(\omega) = 1 - \frac{\omega_p^2}{\omega^2 + i\Gamma\omega} \tag{37}$$

where $\omega_p = 1.37 \times 10^{16}$ rad s$^{-1}$, $\Gamma = 2.73 \times 10^{13}$ rad s$^{-1}$.

As shown in Fig. 2, the CEMD can accurately predict the thermal conductance for a separation distance edge to edge larger than $3R$. For separation distances smaller than $3R$, the CEMD tends to underestimate the thermal conductance, especially for the metal particles. This may be attributed to higher multipoles' contributions which is not included in CEMD. Yet the dipole moment still dominates the radiative heat transfer for a separation distance of $2R$.

## III. RESULTS AND DISCUSSION

In this section, some results of the CEMD are presented. Eq. (28) illustrates that four terms contribute to the radiative heat flux. To show the role played by each of the four terms, the radiative heat flux between different dimers of SiC and Ag nanoparticles is calculated by CEMD. Then, CEMD is applied to investigate the effect of many-body interactions on the NFRHT and the local energy density distribution considering both dielectric and metallic nanoparticles.

### A. The contributions of the four terms to the radiative heat flux

The radiative heat flux between different dimers of SiC and Ag nanoparticles is calculated by CEMD. The first case is between two dimers of SiC particles, the second is between two dimers of Ag particles. The third case is between a dimer of SiC particles and a dimer of Ag particles. In the fourth case, the left dimer is composed of a SiC and a Ag particle while the right is a same dimer but turned upside down. The nanoparticles have a radius of 100nm and in each dimer are separated by $2R$. For simplicity, the temperature of the left dimer is assumed 300K while that of the right is 0K. According to Eq. (27) and (28), the four terms contributing to the spectral radiative heat flux are calculated by

$$P^{\nu\nu'}(\omega) = \sum_{j=1}^{2}\sum_{i=3}^{4} 3\Theta(\omega, T=300K)\mathcal{T}_{ij}^{\nu\nu'}(\omega) \tag{38}$$

where $\nu, \nu' = E, M$, the particles in the left dimer are numbered from 1 to 2 and those of the right are numbered from 3 to 4. Fig. 3 shows the four terms contributing to the spectral radiative heat flux from 0.5 to $5 \times 10^{14}$ rads$^{-1}$ between two dimers separated by $d=0.2\mu$m. Meanwhile, the imaginary parts of the electrical and magnetic $\chi$ [see Eq. (28)] of SiC and Ag are depicted in Fig. 3 (a) and (b), respectively. Note that the electric and magnetic $\chi$ of the SiC and Ag nanoparticles are very close to their electric and magnetic polarizabilities. Overall, the imaginary part of $\chi^E$ of SiC is much larger than its magnetic counterpart. Ag nanoparticle, however, is characterized with large magnetic polarizability.

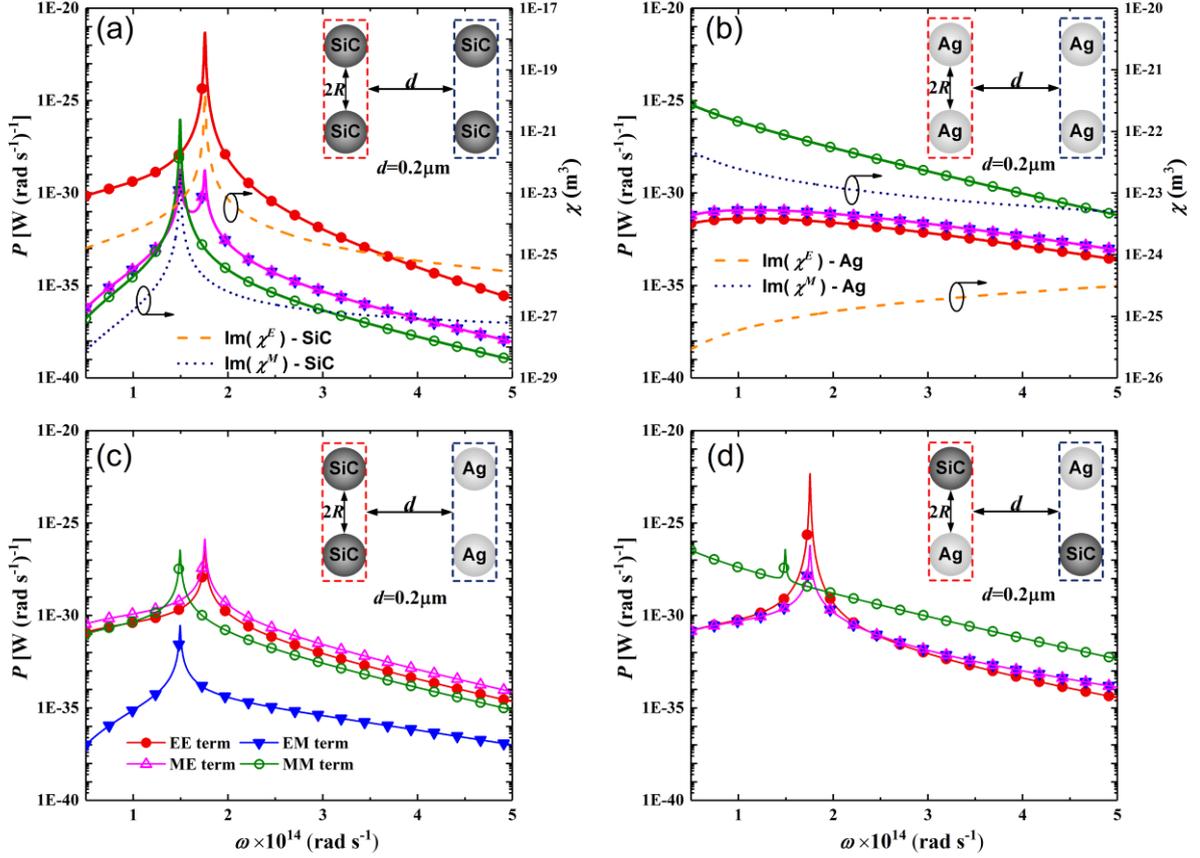

FIG. 3. Contributions of the EE, EM, ME and MM terms [see Eq. (28)] to the spectral radiative heat flux between two dimers of nanoparticles with a separation distance $d=0.2\mu m$. **(a)** two dimers of SiC; **(b)** two dimers of Ag; **(c)** a dimer of SiC and a dimer of Ag; **(d)** a dimer composed of SiC and Ag, and a same dimer but turned upside down. All the particles have a radius of $R=100nm$ and in each dimer are separated by $2R$ edge to edge. The temperature of the left dimer is $T=300K$ while that of the right is $T=0K$. In **(a)** and **(b)**, the imaginary parts of the electrical and magnetic $\chi$ [see Eq. (28)] of SiC and Ag are depicted to the right axis, respectively.

As shown in Fig. 3 (a) for two SiC dimers, the EE term of the spectral radiative heat flux peaks at about $1.756 \times 10^{14} rad \cdot s^{-1}$. This is due to the surface phonon polariton (SPhP) supported by the SiC nanoparticles, as can be illustrated by the peak of $Im(\chi^E)$ for SiC. The MM term peaks at about $1.5 \times 10^{14} rad \cdot s^{-1}$, which can be attributed to the magnetic resonances of the SiC nanoparticles as implied by the peak of $Im(\chi^E)$. The EM and ME terms are quantitatively the same and show two peaks at the frequencies $1.5 \times 10^{14} rad \cdot s^{-1}$ and $1.756 \times 10^{14} rad \cdot s^{-1}$, respectively. As to the EM term, for instance, the first peak at $1.5 \times 10^{14} rad \cdot s^{-1}$ is caused by the large emitting magnetic power due to the magnetic resonance of the emitting dimer. The second peak of the EM term at $1.756 \times 10^{14} rad \cdot s^{-1}$ is caused by the large absorption of the electric power due to the SPhP supported by the absorbing dimer. Yet the magnitude of the two peaks are much smaller than that of the EE term or the MM term. Overall, the EE term is much larger than that of the others. As to the case of two Ag dimers in Fig. 3 (b), the MM term, being much larger than the other three terms, decreases with increasing frequency. The Ag nanoparticles do not support resonances or polaritons in the thermal wavelength range, thus no sharp peaks exist for the four terms of the spectral radiative heat flux. For the third case shown in Fig. 3 (c), the EE and ME terms of the spectral radiative heat flux peak at about $1.756 \times 10^{14}$ rad·s$^{-1}$ due to the SPhP supported by the SiC dimer, whereas the MM and the EM terms peak at $1.5 \times 10^{14}$ rad·s$^{-1}$ due to the magnetic resonances supported by the SiC dimer. In this case, the EM and ME terms are not equal any more. Generally, the ME term is larger than the other terms while the EM term is the smallest. As to the fourth case shown in Fig. 3 (d), each dimer has characteristics of both SiC and Ag nanoparticles, which after mutual interactions leads to a peak at $1.5 \times 10^{14}$ rad·s$^{-1}$ for the MM term and a peaks at $1.756 \times 10^{14}$ rad·s$^{-1}$ for the other three terms. The MM term is larger than the other three terms at frequencies far from that excites SPhP for SiC nanoparticles.

To study further, we calculate the contributions of the four terms to the total radiative heat flux from near- to far-field by

$$P^{\nu\nu'} = \int_0^{+\infty} P^{\nu\nu'}(\omega) \frac{d\omega}{2\pi} \tag{39}$$

where $\nu, \nu' = E, M$. The integration in Eq. (39) is carried over a frequency range from $0.5 \times 10^{14}$ rads$^{-1}$ to $5 \times 10^{14}$ rads$^{-1}$ with sufficient resolution, which is large enough considering the dominant thermal wavelength at $T$=300K.

As shown in Fig. 4 (a), the radiative heat flux between two SiC dimers is dominated by the EE term, which is larger by several orders of magnitude than the other three terms. This is mainly contributed by the SPhP supported by SiC nanoparticles. The MM term, due to the magnetic resonance of SiC nanoparticles, is much larger than the crossed electric and magnetic terms, which is more remarkable in the near-field. For two dimers of Ag nanoparticles shown in Fig. 4 (b), the radiative heat flux is dominated by the MM term. Since the magnetic dipole moments of Ag are much larger than its electric counterpart, the crossed electric and magnetic terms are much larger than the EE term, especially in the far-field. As shown in Fig. 4 (c), the radiative heat flux from the SiC dimer to the Ag dimer is dominated by the ME term, indicating that the radiative heat exchange is mainly contributed by the power dissipation of the magnetic field generated by the electric dipole moments. This is because that the SiC nanoparticles in the emitting dimer supports SPhP while the absorbing Ag dimer is characterized by large power dissipation due to eddy currents. The EM term, on the contrary, is the smallest. The EE and MM terms are comparable from near to far-field. Note that if we reverse the emitting and the absorbing dimers, the radiative heat flux will be dominated by the EM term. For the last case in Fig. 4 (d), the EE term is largest from near to far-field. The MM term is about one order of magnitude larger than the two crossed terms in the near-field, but is much smaller than the two crossed terms in the far-field. In Fig. 4 (a), (b) and (d), the EM term equals the ME term due to the reciprocity of the Green functions and the geometrical symmetry of the dimers. In addition, the EE and MM terms decay approximately as $d^{-4}$ in the near-field with increasing separation distance but decay as $d^{-2}$ in the far-field. The EM and ME terms, however, decay as $d^{-2}$ from near- to far-field.

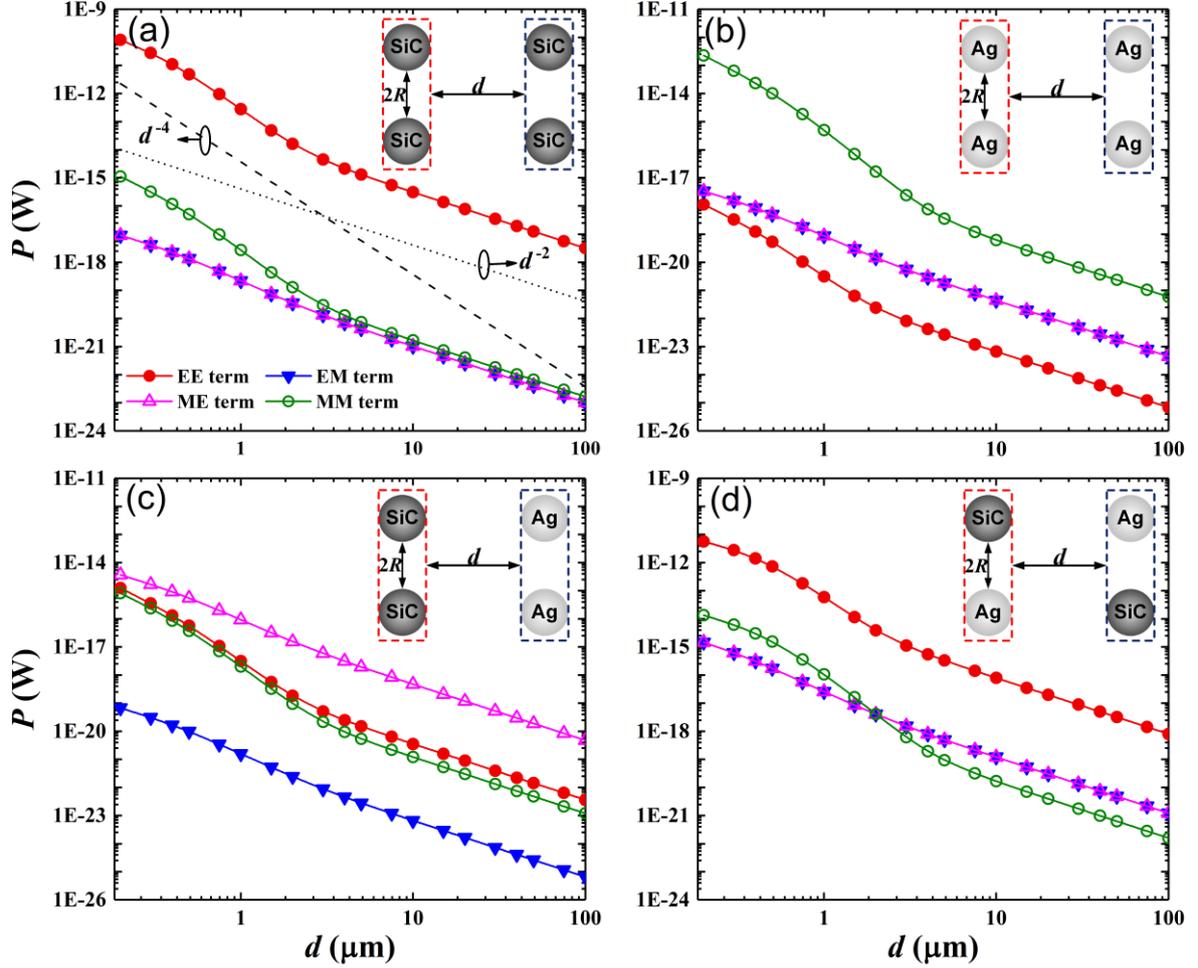

FIG. 4. Contributions of the EE, EM, ME and MM terms [see Eq. (28)] to the total radiative heat flux between two dimers of nanoparticles as a function of separation distance *d*. **(a)** two dimers of SiC; **(b)** two dimers of Ag; **(c)** a dimer of SiC and a dimer of Ag; **(d)** a dimer composed of SiC and Ag, and a same dimer but turned upside down. All the particles have a radius of *R*=100nm and separated by 2*R* edge to edge in each dimer. The left dimer has a temperature of *T*=300K while the right is *T*=0K.

Although we only consider two dimers of nanoparticles, it can be inferred that the radiative heat transfer in a particle system will highly depend on the relative position as well as the composition of the particles. Each of the four terms can dominate the radiative heat transfer. It should be noted that the magnetic polarizabilities of prolonged or oblate metallic nanoparticles (ellipsoids, disks and rings) can largely exceed the electric ones [76]. In that cases, the polarizabilities are anisotropic, which will make the radiative heat transfer in nanoparticle systems more complex. In addition, magnetic polariton can be excited for particles made of magnetic materials, in which case the magnetic dipole moment will play a more important role.

## B. Effect of many-body interactions on NFRHT

Many-body interactions play an important role in the NFRHT as well as the local heat control of nanostructures. The heat flux can be exalted greatly between two SiC nanoparticles when they came into near-field interactions with a third SiC, as shown in Ref. [54] where only the electric dipole moment was considered. CEMD, considering the mutual interactions of the electric and magnetic dipole moments, allows us to investigate the many-body interactions in more cases. For ease of analysis, the normalized heat flux $\varphi$ is defined as

$$\varphi = P_{1\to 2}^3 \big/ P_{1\to 2}^2 \tag{40}$$

where $P_{1\to 2}^3$ denotes the heat flux from particle 1 to particle 2 in a three particle system while $P_{1\to 2}^2$ is the hear flux from 1 to 2 in a two particle system.

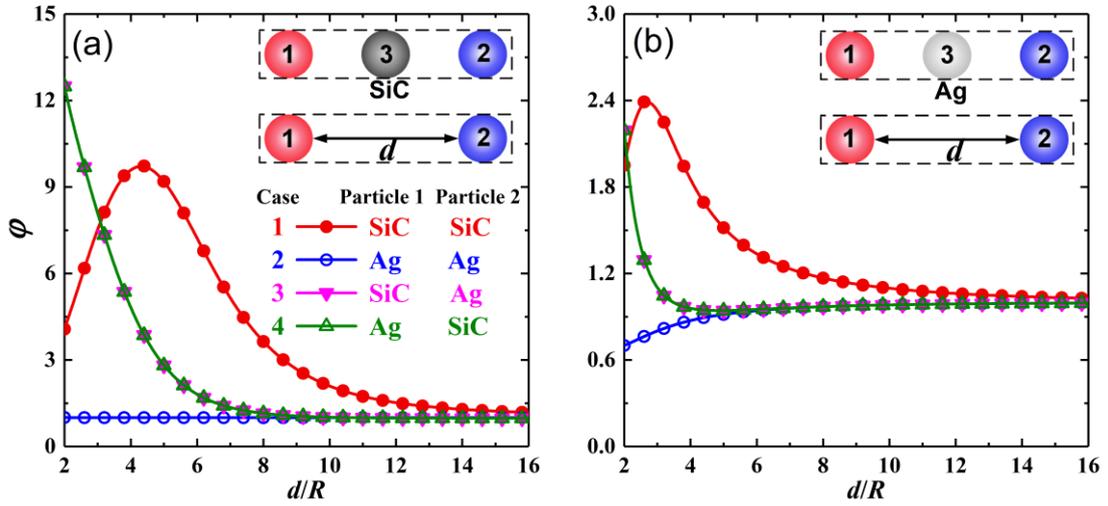

FIG. 5. Normalized radiative heat flux $\varphi$ [see Eq. (40)] from particle 1 to particle 2 with a third particle in the center: **(a)** the third particle is SiC; **(b)** the third particle is Ag. The temperatures of particle 1 and 2 are assumed that $T_1$=300K and $T_2$=0K. $d$ is the separation distance between 1 and 2 edge to edge, $R$=100nm is the radius of the nanoparticles. Four different cases of particle 1 and 2 are considered in (a) and (b).

The normalized heat flux from particle 1 to particle 2 with a third particle placed in the center is shown in Fig. 5. Four cases of particle 1 and 2 are considered, namely SiC to SiC, Ag to Ag, SiC to Ag, and Ag to SiC. The temperatures of particle 1 and 2 are assumed to be 300K and 0K, respectively. In Fig. 5 (a) the third particle is SiC. As shown in Fig. 5 (a), the radiative heat flux between two SiC nanoparticles is greatly enhanced by a third SiC nanoparticle, which is similar to that observed in Ref. [54]. The normalized heat flux $\varphi$ reaches a maximum of about 10 when the separation distance between particle 1 and 2 is about $5R$. With increasing separation distance, however, the heat flux enhancement becomes less obvious. For the case of two Ag nanoparticles, on the contrary, the effect on the heat flux is negligible when a SiC particle is placed in the center. For the third case where the emitter is SiC and the absorber is Ag, the heat flux can be enhanced greatly when is the separation distance between 1 and 2 is smaller than 6*R*. Yet, *φ* decreases quickly with increasing distance and approaches unity when *d* is larger than 8*R*. The fourth case yields the same results to that of the third case. The same four cases are considered in Fig. 5 (b) except that the third particle is Ag. As to the first case, the Ag particle, although not resonant in the thermal wavelength range, can enhance radiative heat flux between SiC and SiC. However, the normalized heat flux is much smaller compared to the first case in Fig. 5 (a), and the maximum of *φ* shifts to smaller separation distance. As to the second case, the Ag nanoparticle will decrease the heat flux between two Ag nanoparticles when the separation distance is smaller than 4*R* but have no effect for larger distances. Note that if higher multipoles are included in this case, the situation may be different. For the third and fourth cases, similar phenomena can be observed as in Fig. 5 (a) but the effect on the heat flux by the Ag nanoparticle is much weaker.

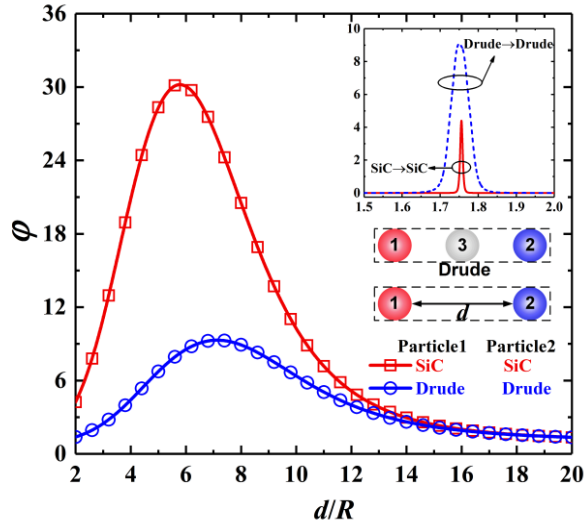

FIG. 6. Normalized radiative heat flux *φ* [see Eq. (40)] from particle 1 to particle 2 with a Drude particle in the center. For the Drude particle, $\omega_p$ is set as $3.042\times10^{14}$rad·s$^{-1}$ and $\Gamma$ is set as $0.01\omega_p$ so that the Drude particle supports SPP at about the same frequency of SiC supporting SPhP. *d* is the separation distance between 1 and 2 edge to edge, *R*=100nm is the radius of the particles. The temperatures of particle 1 and 2 are assumed that $T_1$=300K and $T_2$=0K. The upper right inset shows the radiative heat flux ($\times10^{16}$) from 1 to 2 without the third particle in a spectral range from $1.5\times10^{14}$rad·s$^{-1}$ to $2.0\times10^{14}$rad·s$^{-1}$.

For nanoparticles with a Drude-like dielectric function, localized surface plasmon polariton (SPP) can be excited at a frequency of about $\omega_p/\sqrt{3}$ [77]. SPP of Ag nanoparticles is excited in the ultraviolet range, which is far from the characteristic thermal wavelengths considered. However, some materials like the doped semiconductors can support SPP in the infrared range. To see if particles supporting SPP have similar phenomena as SiC, a Drude particle is introduced, for which $\omega_p$=$3.042\times10^{14}$rad·s$^{-1}$ and $\Gamma$=$0.01\omega_p$. Thus, the Drude particle supports SPP at about $1.756\times10^{14}$rad·s$^{-1}$, the same frequency as SiC nanoparticles supporting SPhP. The normalized heat flux from particle 1 to particle 2 with a Drude particle in the center is shown in Fig. 6. Two cases are considered, i.e. from SiC to SiC and from Drude to Drude. As illustrated in Fig. 6, the radiative heat flux can be enhanced by several tens when a Drude particle is placed in the center between two SiC nanoparticles, which is much larger than the case where a SiC is placed in the center. This is because the SPP supported by the Drude particle is much stronger than the SPhP supported by SiC, which can be reflected by the heat flux near the resonance frequencies as shown in the upper right inset of Fig. 6. In addition, the maximum of $\varphi$ shifts to larger separation distance. This indicates that SPP and SPhP can be coupled to enhance near-field radiative heat transfer, and the heat flux can be tuned by the resonance strength. As to the case of two Drude particles, the heat flux can also be enhanced by a third Drude particle placed in the center. In this case, however, the maximum of $\varphi$ occurs for a larger separation distance of about $7.5R$.

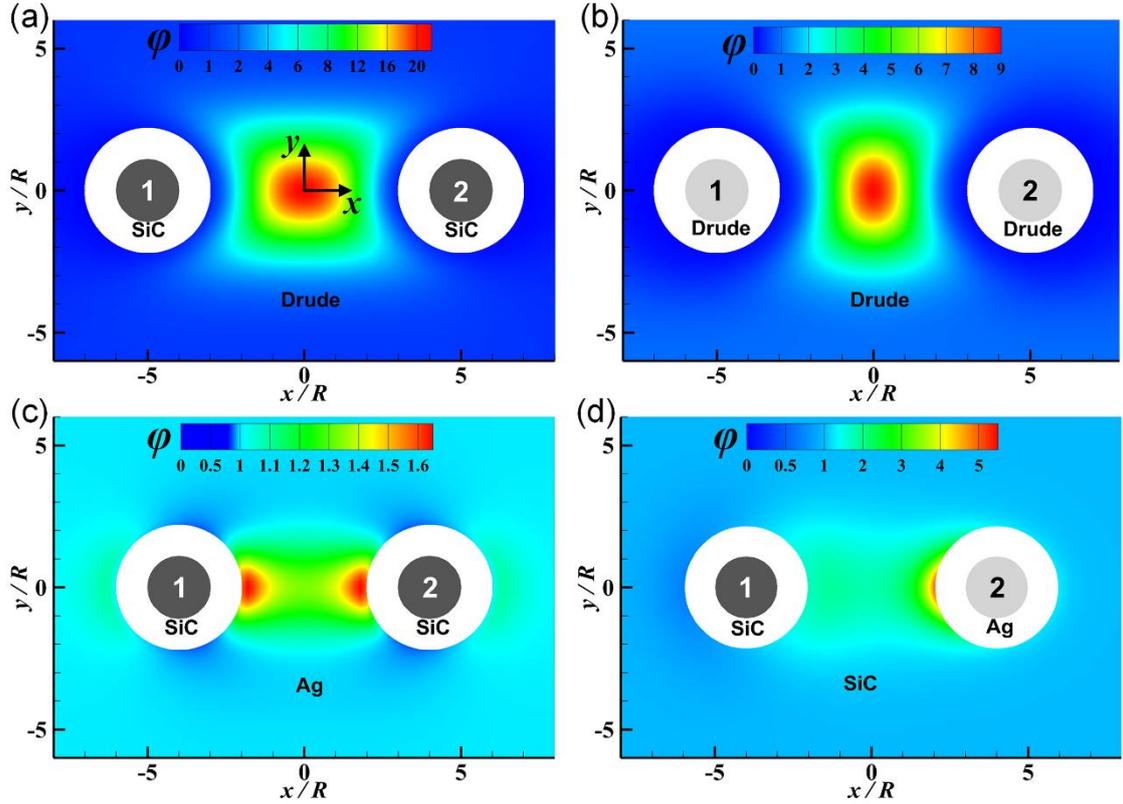

FIG. 7. Normalized radiative heat flux $\varphi$ [see Eq. (40)] from particle 1 to particle 2 with respect to the position of a third particle. **(a)** two SiC particles separated by $8R$ edge to edge, the third particle is a Drude particle; **(b)** two Drude particles separated by $8R$ with a third Drude particle; **(c)** two SiC separated by $6R$, the third particle is a Ag; **(d)** a SiC and a Ag separated by $6R$, the third particle is a SiC. For the Drude particle, $\omega_p=3.042\times10^{14}$rad·s$^{-1}$ and $\Gamma=0.01\omega_p$ so that the Drude particle supports SPP at about the same frequency of SiC supporting SPhP. The temperatures of particle 1 and 2 are $T_1$=300K and $T_2$=0K. The radius of all the particles is $R$=100nm. The white circle denotes the area where the third particle cannot be placed.

To give more details, the normalized radiative heat flux from particle 1 to particle 2 with respect to the position of a third particle in 2D space is demonstrated in Fig. 7. According to the results analyzed above, we choose four cases, namely (a) between two SiC particles with a Drude particle, (b) between two Drude particles with a Drude particle, (c) between two SiC particles with a Ag particle, and (d) between a SiC and a Ag particle with a SiC particle. For cases (a) and (b), particle 1 and 2 are separated by 8*R* edge to edge, whereas for cases (c) and (d), particle 1 and 2 are separated 6*R*. As to the cases in Fig. 7 (a) and (b), the enhancement of the heat flux is largest at the center between particle 1 and 2, which is similar to that observed in Ref. [54]. Yet, the third Drude particle can also decrease the heat flux when it is close to either the emitter or the absorber, which is clearer for case (b). As the third particle gets far away from particle 1 and 2, its effect on the radiative heat flux becomes negligible. For the case in Fig. 7 (c), the Ag nanoparticle can increase the heat flux when it is aligned with the two SiC nanoparticles. The normalized heat flux, although much smaller than the first two cases, is larger when Ag is closer to either the emitter or the absorber. For the case in Fig. 7 (d), the SiC nanoparticle can increase the radiative heat flux from SiC to Ag in a relatively larger region. And $\varphi$ is much larger when the SiC particle is closer to the front of Ag, the non-resonant particle. Note that if we reverse particle 1 and 2 in case (d), the region map will also be reversed.

## C. Effect of many-body interactions on the local energy density distribution

The effect of many-body interactions on the local energy density distribution has not been studied. According to results of the last sub-section, four cases are chosen. The first case, as shown in Fig. 8 (a), is two SiC nanoparticles separated by 8*R* edge to edge. In cases (b) and (c), a Drude and a Ag particle is placed in the center of the two SiC particles. Case (d) is similar to case (b) but the Drude particle is closely in front of the right SiC particle. The left SiC has a temperature of *T*=300K while the other particles is 0K. The frequency is set as $1.756 \times 10^{14}$ rad·s$^{-1}$, the local energy density is calculated by Eqs. (30) and (31).

As shown in Fig. 8 (a), the energy density of the electromagnetic wave from the left SiC nanoparticle decreases radially with increasing distance. However, the local energy density increases dramatically around the right SiC nanoparticle, which can be attributed to the existence of surface waves of SPhP excited by the incident electromagnetic waves from the left SiC particle. As shown in Fig. 8 (b), the local energy density is further increased when a Drude particle supporting SPP is placed in the center, which explains why the radiative heat transfer is enhanced greatly when the Drude particle in placed in the center of two SiC particles. As to the case in Fig. 8 (c) where a non-resonant Ag nanoparticle is placed in the center, the local energy density distribution shows little difference from case (a). For the case in Fig. 8 (d) where a Drude particle is placed closely in front of the right SiC nanoparticle, the local energy density around the Drude and the right SiC particle decreases dramatically compared to that of case (b). Note the local energy density around the right SiC is even smaller than that shown in Fig. 8 (a). This agrees with the phenomena observed in Fig. 7 (b), i.e. the radiative heat flux is decreased when a third resonant particle is placed close to the absorbing particle. However, it must be emphasized that in this case higher multipoles may play an important role.

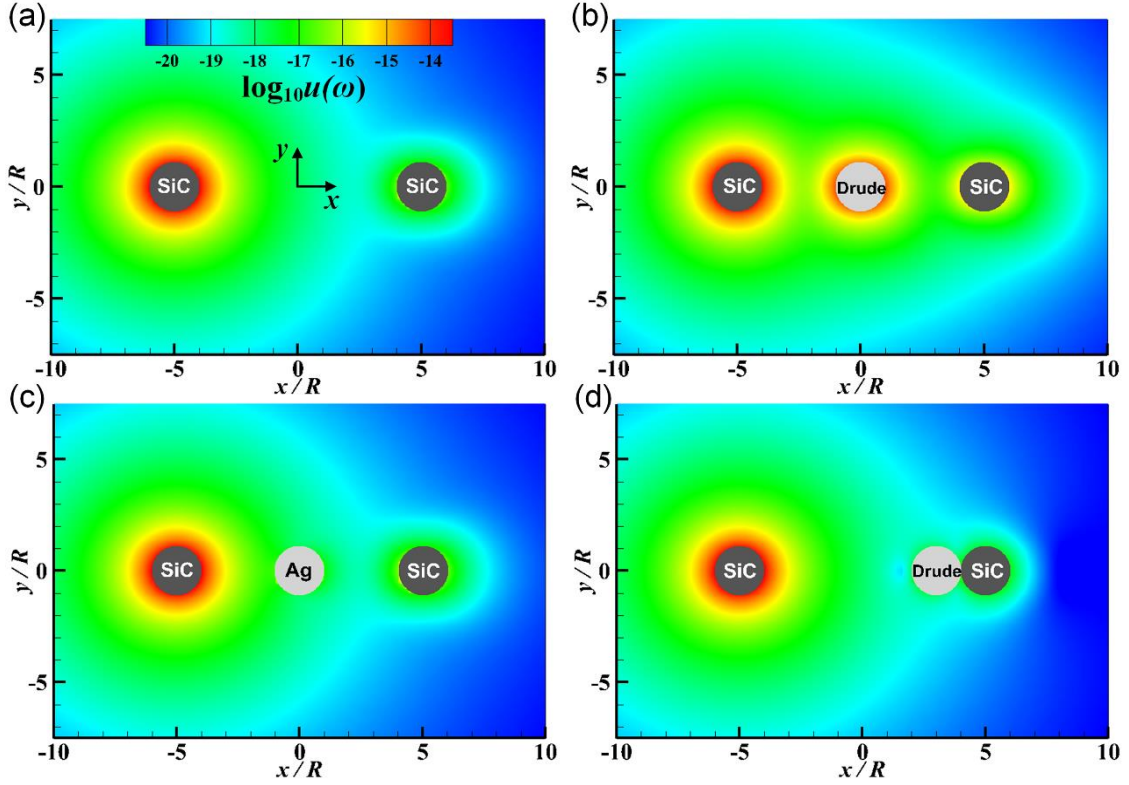

FIG. 8. Local energy density $u(\omega)$ in vacuum at the frequency $1.756 \times 10^{14}$ rad·s$^{-1}$, **(a)** two SiC nanoparticles; **(b)** two SiC nanoparticles with a Drude particle in the center; **(c)** two SiC nanoparticles with a Ag particle in the center; **(d)** similar as (b) but the Drude particle is closely in front of the second SiC particle. The left SiC has a temperature of 300K while the other particles are 0K. The radius of the particles is $R$=100nm. For the Drude particle, $\omega_p = 3.042 \times 10^{14}$ rad·s$^{-1}$ and $\Gamma = 0.01\omega_p$ so that the Drude particle supports SPP at about the same frequency of SiC supporting SPhP.

## IV. CONCLUSIONS

In conclusion, we have developed the coupled electric and magnetic dipole (CEMD) approach for the radiative heat transfer in nanoparticle systems. Combined with the fluctuation-dissipation theorem, the CEMD solves the electric and magnetic dipole moments of the nanoparticles in mutual interactions. The radiative heat flux and the local energy density are deduced once the Green's functions of the particle system are obtained. Four terms, namely the electric-electric term, the electric-magnetic term, the magnetic-electric term and the magnetic-magnetic term, contribute to the radiative heat flux and the local energy density. With CEMD, we have studied the radiative heat transfer between various dimers of SiC and Ag nanoparticles. It was found that the radiative heat transfer highly depends on the relative position and the composition of the particle system. Each of the four terms can dominate the radiative heat transfer. In addition, the CEMD was applied to study of the effect of many-body interactions on the near-field radiative heat transfer (NFRHT) and the local energy density distribution considering both dielectric and metallic nanoparticles. Provided that the particles are in coupled resonances, the near-field radiative heat flux and local energy density can be greatly increased. SPP and SPhP can be coupled to enhance the radiative heat flux, and the enhanced radiative heat flux can be further modulated by the resonance strength.

The CEMD can find applications for the study of the radiative heat transfer through random nanoscale systems, for example, densely packed nanoparticle beds and nanofluids. In such systems, the role played by NFRHT needs to be elucidated and multiscale simulations [78,79] should be considered for the coupled heat transfer by conduction, convection and radiation. Much room is left for the many-body radiative heat transfer theory. The many-body radiative heat transfer theory considering the interactions with infinite planar geometries needs to be developed, which will be useful to study the NFRHT for nanoparticles placed in nanochannels [80,81]. To go beyond the dipole approximation, the many-body radiative heat transfer theory should include higher electric and magnetic multipoles for smaller separation distances and consider multiple larger finite sized objects. In addition, the magnetic dipole moment considered in this work is caused by the eddy current in the particle. Yet the CEMD can be easily extended to magnetic materials.

**ACKNOWLEDGMENTS**

We gratefully acknowledge the supports by the National Natural Science Foundation of China (Nos. 51336002, 51421063). JMZ also acknowledges the support by the Fundamental Research Funds for the Central Universities (Grant No. HIT.BRETIII.201415). JD sincerely thanks A. Narayanaswamy and B. Czapla for providing the codes of the near-field radiative heat transfer between two spheres. In addition, we are grateful to the two anonymous reviewers who provided kind and helpful suggestions.